\documentclass[prb,reprint,superscriptaddress,citeautoscript]{revtex4-1}

\usepackage[applemac]{inputenc}
\usepackage[T1]{fontenc}
\usepackage[american]{babel}
\usepackage[scaled]{helvet}
\usepackage{graphicx,amssymb,bm,mathtools,textcomp,courier,multirow,bigdelim,color}
\usepackage[normalem]{ulem}	

\newcommand{\op}[2]{\left.{\left| #1\right\rangle\left\langle #2\right|}\right.}

\newcommand{\ket}[1]{\left|\left. #1 \right\rangle\right.}

\newcommand{\cre}[1]{c^{\dagger}_{#1}}
\newcommand{\ann}[1]{c_{#1}}

\newcommand{\figureshortname}{Fig.}
\newcommand{\equationshortname}{Eq.}
\newcommand{\tableshortname}{Tab.}

\newcommand{\eref}[1]{\equationshortname~\eqref{#1}}
\newcommand{\sref}[1]{Sec.~\ref{#1}}
\newcommand{\aref}[1]{Appx.~\ref{#1}}
\newcommand{\cref}[1]{Chapter~\ref{#1}}
\newcommand{\fref}[1]{\figureshortname~\ref{#1}}
\newcommand{\tref}[1]{\tableshortname~\ref{#1}}
\newcommand{\rcite}[1]{Ref.~[\onlinecite{#1}]}

\graphicspath{{Pictures/}}

\begin{document}
\def\sectionautorefname{Sec.}

\title{Two-Qubit Pulse Gate for the Three-Electron Double Quantum Dot Qubit}

\author{Sebastian Mehl}
\email{s.mehl@fz-juelich.de}
\affiliation{Peter Gr\"unberg Institute (PGI-2), Forschungszentrum J\"ulich, D-52425 J\"ulich, Germany}
\affiliation{JARA-Institute for Quantum Information, RWTH Aachen University, D-52056 Aachen, Germany}

\date{\today}

\begin{abstract}
The three-electron configuration of gate-defined double quantum dots encodes a promising qubit for quantum information processing. I propose a two-qubit entangling gate using a pulse-gated manipulation procedure. The requirements for high-fidelity entangling operations are equivalent to the requirements for the pulse-gated single-qubit manipulations that have been successfully realized for Si QDs. This two-qubit gate completes the universal set of all-pulse-gated operations for the three-electron double-dot qubit and paves the way for a scalable setup to achieve quantum computation.
\end{abstract}

\maketitle

\section{Introduction}
The name hybrid qubit (HQ) was coined for the qubit encoded in a three-electron configuration on a gate-defined double quantum dot (DQD) \cite{shi2012,koh2012}. The HQ is a spin qubit in its idle configuration, but it is a charge qubit during the manipulation procedure. Recently, impressive progress was made for the single-qubit control of a HQ in Si \cite{shi2014,kim2014}. It was argued that single-qubit gates were implemented, whose fidelities exceed $85~\%$ for X rotations and $94~\%$ for Z rotations \cite{kim2014}. 
These manipulations rely on the transfer of one electron between quantum dots (QDs) \cite{koh2012,shi2014,kim2014}. Subnanosecond gate pulses were successfully applied to transfer the third electron between singly occupied QDs.

\rcite{shi2012} suggested two-qubit gates between HQs with similar methods to these for three-electron spin qubits that are defined at three QDs \cite{divincenzo2000,fong2011}. The coupling strength between neighboring QDs is tuned in a multi-step sequence, while this entangling gate for HQs requires control over the spin-dependent tunnel couplings. A more realistic approach to realize two-qubit entangling gates for HQs uses electrostatic couplings between the HQs \cite{koh2012}. If the charge configuration of one HQ is changed, then Coulomb interactions modify the electric field at the position of the other HQ. Note the equivalent construction for a controlled phase gate ($\text{CPHASE}$) for singlet-triplet qubits in two-electron DQDs \cite{hanson2007-1}.

Using Coulomb interactions for entangling operations can be critical. Even though electrostatic couplings are long-ranged, they are generally weak and they are strongly disturbed by charge noise \cite{shulman2012}. I propose an alternative two-qubit gate. Two HQs in close proximity enable the transfer of electrons. The two-qubit gate that is constructed works similarly to the pulse-gated single-qubit manipulations. It requires fast control of the charge configurations on the four QDs through subnanosecond pulse times at gates close to the QDs. A two-qubit manipulation scheme of the same principle as for the single-qubit gates is highly promising because single-qubit pulse gates have been implemented with great success \cite{shi2014,kim2014}.

The central requirement of the entangling operation is the tuning of one two-qubit state to a degeneracy point with one leakage state 
(called $\ket{E}$).
The qubit states are $\ket{1}$ and $\ket{0}$, while the subscripts $L$ and $R$ describe the physical positions of the HQs. Specifically, when the state $\ket{0_L0_R}$ is degenerate with $\ket{E}$ then $\ket{0_L0_R}$ can pick up a nontrivial phase, while all the other two-qubit states evolve trivially. Note that a similar construction for an entangling operation \cite{strauch2003} has been implemented with impressive fidelities \cite{dicarlo2009,dicarlo2010,barends2014} for superconducting qubits. The couplings to other leakage states must be avoided during the operation. I propose a two-step procedure. First, $\ket{1_L1_R}$ and $\ket{0_L1_R}$ are tuned away from the initial charge configuration to protect these states from leakage. $\ket{1_L0_R}$ and $\ket{0_L0_R}$ remain unchanged at the same time. One has then reached the readout regime of the second HQ. The second part of the tuning procedure corrects the passage of $\ket{1_L0_R}$ through the anticrossing with $\ket{E}$, at a point where $\ket{1_L0_R}$ is degenerate with 
another leakage state (called $\ket{L}$). I call this anticrossing degenerate Landau-Zener crossing (DLZC) because the passage through this anticrossing is described by a generalization of the Landau-Zener model \cite{usuki1997,vasilev2007}.

I focus on pulse-gated entangling operations for HQs in gate-defined Si QDs. Even though the entangling operation is not specifically related to the material and the qubit design, gate-defined Si QDs are the first candidate where the two-qubit pulse gate might be implemented because Si QDs were used for single-qubit pulse gates \cite{shi2014,kim2014}. I discuss therefore specifically the noise sources that are dominant for experiments involving gate-defined Si QDs. The described two-qubit pulse gates can be directly implemented with the existing methods of the single-qubit pulse gates. It will turn out that high-fidelity two-qubit entangling operations require low charge noise.

The organization of this paper is as follows. \sref{sec:Setup} introduces the model to describe a pair of three-electron DQDs. \sref{sec:PGate} constructs the two-qubit gate. \sref{sec:Noise} discusses the noise properties of the entangling operation, and \sref{sec:Concl} summarizes all the results.

\section{Setup
\label{sec:Setup}}
\begin{figure}[htb]
\centering
\includegraphics[width=0.4\textwidth]{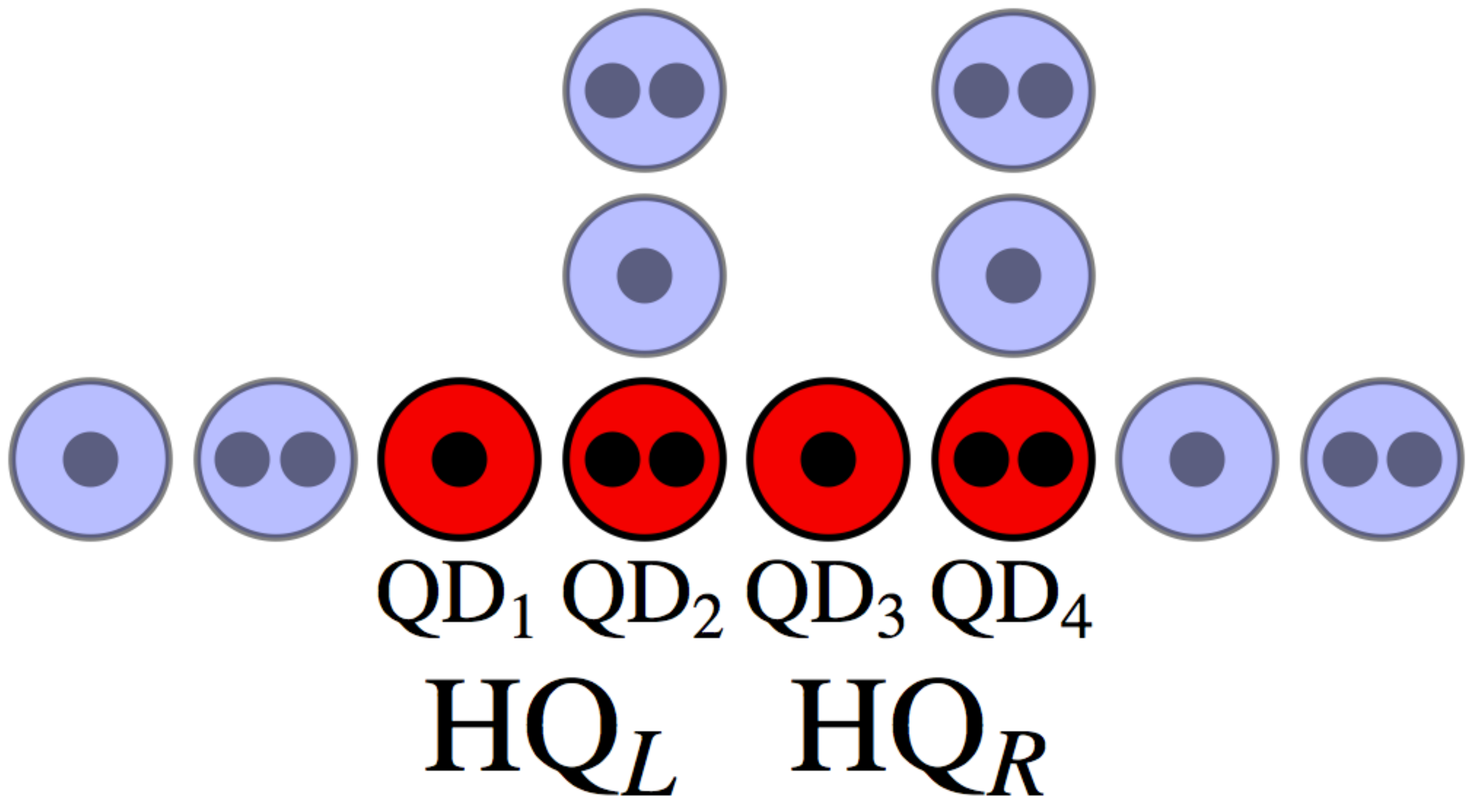}
\caption{Array of QDs that is used to define and couple HQs. The four red QDs encode two HQs, and they are labeled by $\text{HQ}_{L}$ and $\text{HQ}_R$. Black dots represent electrons. The charge configurations are labeled by the electron numbers $\left(n_{\text{QD}_1},n_{\text{QD}_2},n_{\text{QD}_3},n_{\text{QD}_4}\right)$. $\left(1,2,1,2\right)$ is the idle configuration. Applying voltages to gates close to the QDs provides universal single-qubit control and realizes a $\text{CPHASE}$ gate by the transfer of single electrons between the QDs. The gate protocols to achieve quantum computation are described in the text. The encoding scheme can be scaled up trivially, as shown by the blue QDs.
\label{fig:1}}
\end{figure}

I consider an array of four QDs, which are labeled by $\text{QD}_{1}$-$\text{QD}_{4}$ (see \fref{fig:1}). One qubit is encoded using a three-electron configuration on two QDs. $\text{QD}_{1}$ and $\text{QD}_{2}$ encode $\text{HQ}_L$, and $\text{QD}_{3}$ and $\text{QD}_{4}$ encode $\text{HQ}_R$. The system is described by a Hubbard model, which includes two orbital states at each QD. The transfer of electrons between neighboring QDs is possible but weak, unless the system is biased using electric gates. 
It might be desirable to apply a large global magnetic field, which separates states of different $s_z$ energetically. 
Generally, such a global magnetic field is not needed for the pulse-gated entangling operation because the electron transfer between QDs is spin conserving for weak spin-orbit interactions (as for all Si heterostructures). Also nuclear spin noise only introduces a very small spin-flip probability \cite{zwanenburg2013}. Nevertheless, a global magnetic field still reduces the influence of the remaining nuclear spin noise.

The $S=\frac{1}{2}$, $s_z=\frac{1}{2}$ spin subspace of three electrons is two dimensional, and it encodes a qubit \cite{divincenzo2000}.
The single-qubit states for $\text{HQ}_L$ are 
$\ket{1_{L}}=
\sqrt{\frac{2}{3}}\ket{\downarrow T_+}-
\sqrt{\frac{1}{3}}\ket{\uparrow T_0}$ and 
$\ket{0_{L}}=
\ket{\uparrow S}$. The first entry in the state notation labels electrons at $\text{QD}_1$, and the second entry labels electrons at $\text{QD}_2$. $\text{QD}_{1}$ is singly occupied, but two electrons are paired at $\text{QD}_2$.
$\ket{S}=\cre{i\uparrow}\cre{i\downarrow}\ket{0}$ is the two-electron singlet state at $\text{QD}_i$, 
$\ket{T_{+}}=\cre{i\uparrow}\cre{\overline{i}\uparrow}\ket{0}$, 
$\ket{T_{0}}=\frac{1}{\sqrt{2}}\left(\cre{i\uparrow}\cre{\overline{i}\downarrow}+\cre{i\downarrow}\cre{\overline{i}\uparrow}\right)\ket{0}$, and $\ket{T_{-}}=\cre{i\downarrow}\cre{\overline{i}\downarrow}\ket{0}$ are triplet states at $\text{QD}_i$. $c_{i\sigma}^{\left(\dagger\right)}$ is the (creation) annihilation operator of one electron in state $\ket{i}$ of $\text{QD}_i$ with spin $\sigma$, $\ket{i}$ and $\ket{\overline{i}}$ are the ground state and the first excited state at $\text{QD}_i$,\footnote{
Note that $\protect{\ket{\overline{i}}}$ can be an orbital excited state or a valley excited state in Si. $\protect{\ket{2}}$ and $\protect{\ket{\overline{2}}}$ determine the energy difference between $\protect{\ket{0_L}}$ and $\protect{\ket{1_L}}$. The described two-qubit gate relies on a larger singlet-triplet energy difference for the two-electron configuration at $\text{QD}_1$ compared to $\text{QD}_2$. Equivalent discussions hold for $\text{HQ}_R$. In contrast, GaAs QDs lack valley excited states; one can realize the same entangling gate using one large QD and one small QD. Then, the energy difference between the two-electron singlet and the two-electron triplet depends on the confining strength of the wave functions.
} 
and $\ket{0}$ is the vacuum state. Similar considerations hold for $\text{HQ}_R$, where $\text{QD}_3$ is singly occupied and $\text{QD}_4$ is filled with two electrons. It is assumed that a two-electron triplet at $\text{QD}_1$ or at $\text{QD}_3$ is strongly unfavored compared to a two-electron triplet at $\text{QD}_2$ or at $\text{QD}_4$. These conditions were fulfilled for the HQs in \rcite{shi2014} and \rcite{kim2014}.

The energy $E_0=0$ is assigned to $\ket{0_L0_R}$ in $\left(1,2,1,2\right)$. $\ket{1_L0_R}$, $\ket{0_L1_R}$, and $\ket{1_L1_R}$ are higher in energy by $\Omega_{L}$, $\Omega_{R}$, and $\Omega_{L}+\Omega_R$. The excited states $\ket{1_L}$ and $\ket{1_R}$ involve a triplet on a doubly occupied QD that is higher in energy than the singlet configurations of $\ket{0_L}$ and $\ket{0_R}$. Single-qubit gates are not the focus of this work, but I briefly review: all single-qubit gates are applicable through evolutions under $\sigma_x^{L}$, $\sigma_z^{L}$, $\sigma_x^{R}$, and $\sigma_z^{R}$. $\sigma_x=\op{1}{0}+\op{0}{1}$ and $\sigma_z=\op{1}{1}-\op{0}{0}$ are the Pauli operators on the corresponding qubit subspace. They are applied by transferring one electron from $\text{QD}_2$ to $\text{QD}_1$ for $\text{HQ}_L$ (and $\text{QD}_4$ to $\text{QD}_3$ for $\text{HQ}_R$). Depending on the pulse profile, pure phase evolutions (described by the operators $\sigma_z^{L}$ and $\sigma_z^{R}$) or spin flips (described by the operators $\sigma_x^{L}$ and $\sigma_x^{R}$) are created \cite{koh2012,shi2014,kim2014}.

\section{Two-Qubit Pulse Gate
\label{sec:PGate}}
Two-qubit operations are constructed using the transfer of electrons between neighboring QDs. The charge transfer between $\left(1,2,1,2\right)$ and $\left(1,2,2,1\right)$ is described by $\mathcal{H}_{34}=
\tau_1\sum_{\sigma\in\left\{\uparrow,\downarrow\right\}}\left(\cre{3\sigma}\ann{4\sigma}+\text{H.c.}\right)+
\tau_2\sum_{\sigma\in\left\{\uparrow,\downarrow\right\}}\left(\cre{3\sigma}\ann{\overline{4}\sigma}+\text{H.c.}\right)$, where $\tau_1$, $\tau_2$ are tunnel couplings between states from neighboring QDs, and H.c. labels the Hermitian conjugate of the preceding term. $\epsilon_{43}= eV_4-eV_3$ describes the transfer of electrons through voltages applied at gates close to $\text{QD}_3$ and $\text{QD}_4$. Lowering the potential at $\text{QD}_3$ compared to $\text{QD}_4$ favors $\left(1,2,2,1\right)$ ($\epsilon_{43}>0$), but $\left(1,2,1,2\right)$ is favored for the opposite case ($\epsilon_{43}<0$). $\left(1,2,1,2\right)$ and $\left(1,2,2,1\right)$ have identical energies at $\epsilon_{43}=\Delta_{43}>\Omega_L,\Omega_R$. Similar considerations hold for the manipulation between $\left(1,2,1,2\right)$ and $\left(1,1,2,2\right)$, which is described by $\epsilon_{23}=eV_2-eV_3$ and $\mathcal{H}_{23}=
\tau_3\sum_{\sigma\in\left\{\uparrow,\downarrow\right\}}\left(\cre{2\sigma}\ann{3\sigma}+\text{H.c.}\right)+
\tau_4\sum_{\sigma\in\left\{\uparrow,\downarrow\right\}}\left(\cre{\overline{2}\sigma}\ann{3\sigma}+\text{H.c.}\right)$. $\left(1,2,1,2\right)$ and $\left(1,1,2,2\right)$ have identical energies at $\epsilon_{23}=\Delta_{23}>\Omega_L,\Omega_R$.

Note that electrostatic couplings between the states of different charge configurations are neglected in this discussion. \rcite{koh2012} argued that the Coulomb interaction can introduce energy shifts of $\gtrsim 0.1~\mu\text{eV}$, reaching the magnitudes of the orbital energies (typically $0.1-10~\mu\text{eV}$). Coulomb interactions modify the state energies of different charge configurations [we consider only $\left(1,2,1,2\right)$, $\left(1,2,2,1\right)$, and $\left(1,1,2,2\right)$]. These modifications do not influence the operation principle of the entangling gate because only a two-qubit system with a state degeneracy with one leakage state is required. The Coulomb interactions can be introduced by a shift of the positions of the state degeneracies between different charge configurations.

\begin{figure}[h!]
\centering
\includegraphics[width=0.49\textwidth]{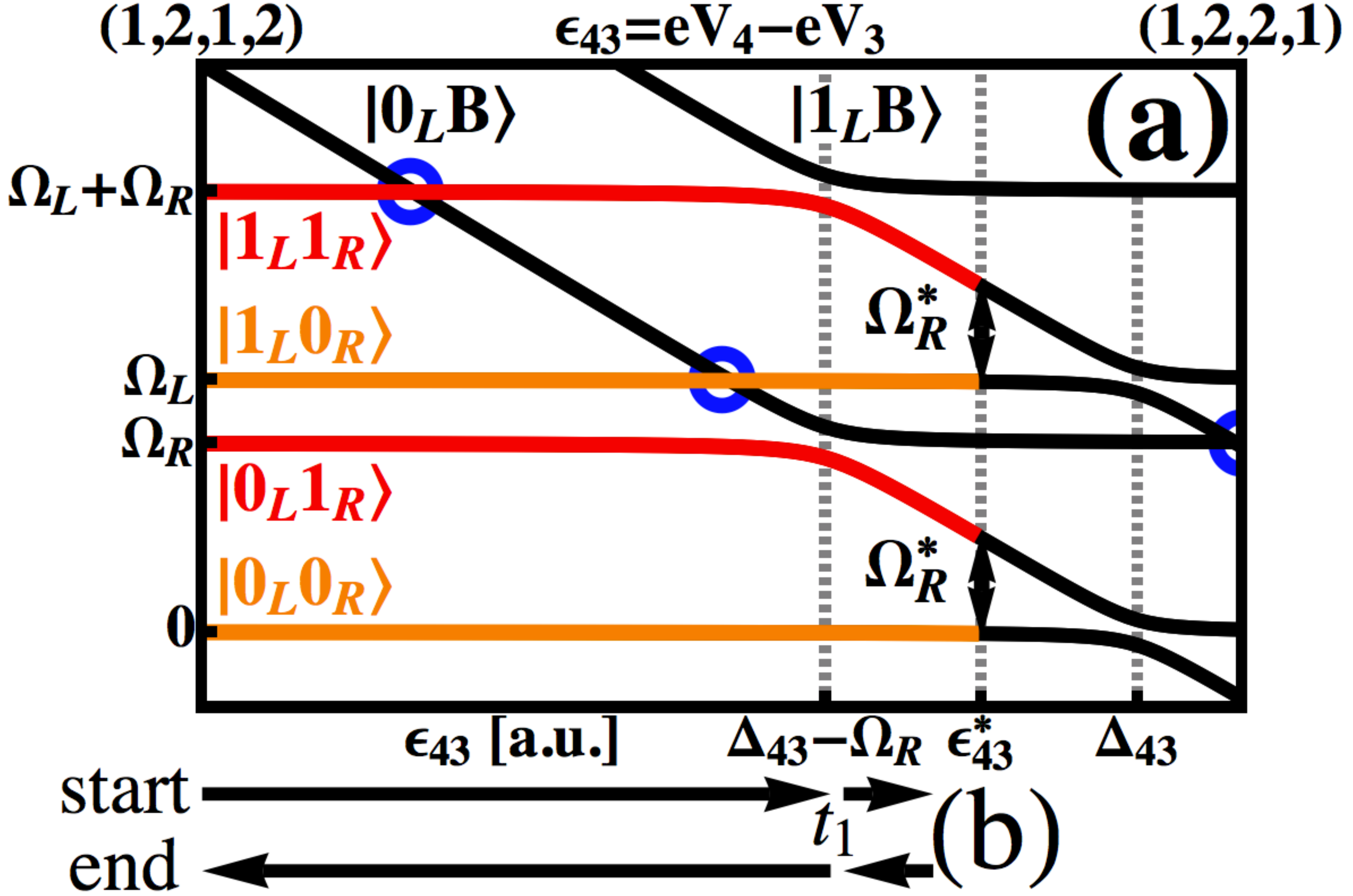}

\bigskip
\includegraphics[width=0.49\textwidth]{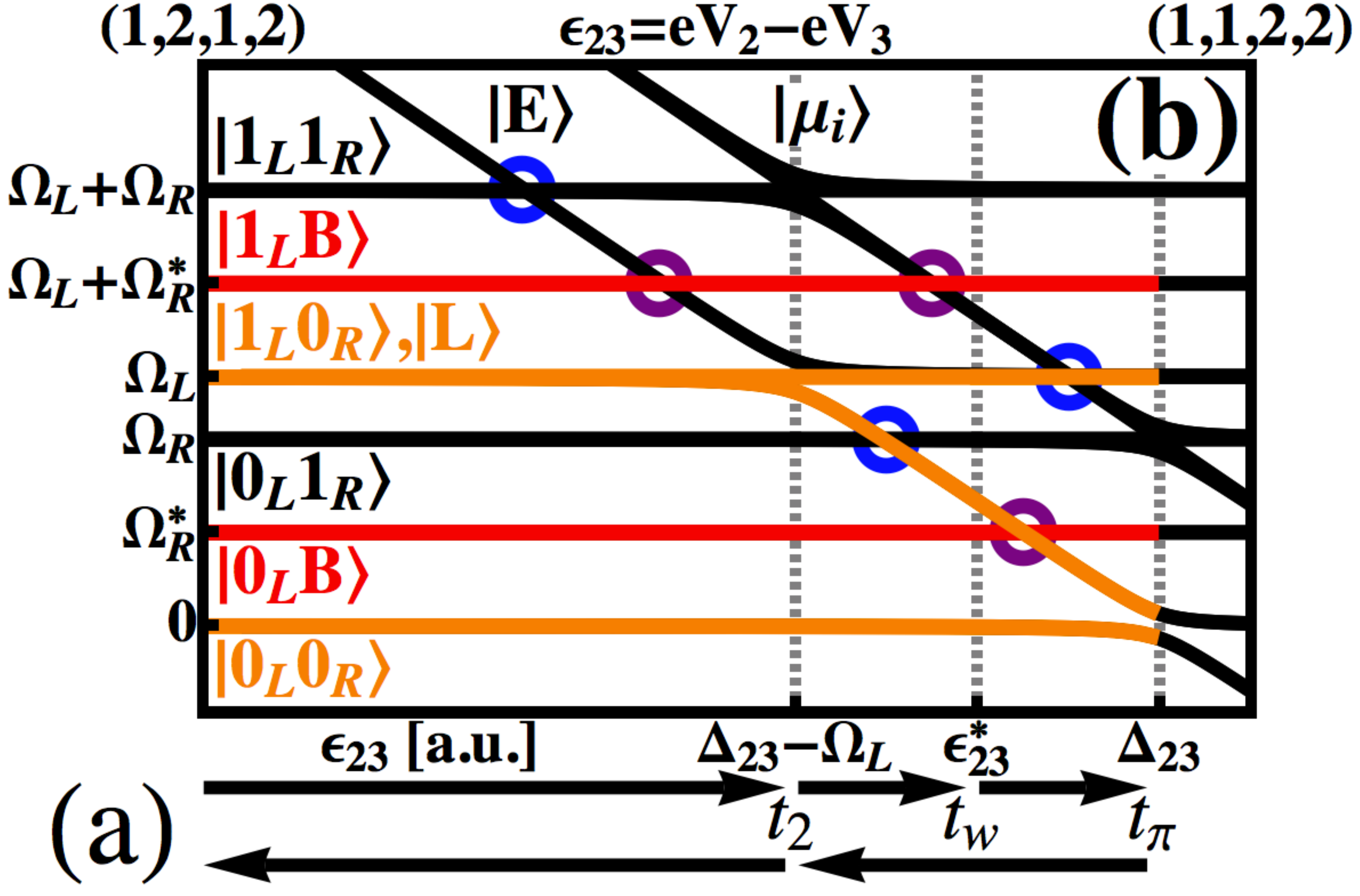}
\caption{Energy diagram of two coupled HQs with $s_z=1$ in $\left(1,2,1,2\right)$, $\left(1,2,2,1\right)$, and $\left(1,1,2,2\right)$. The red and the orange lines describe computational basis, and the black lines are leakage states. $\left(1,2,1,2\right)$ is favored without external bias. (a) shows the pulsing towards $\left(1,2,2,1\right)$, which is modeled through $\epsilon_{43}=eV_4-eV_3$ describing the potentials at $\text{QD}_3$ and $\text{QD}_4$. The states $\ket{1_L1_R}$ and $\ket{1_L B}$ as well as $\ket{0_L1_R}$ and $\ket{0_L B}$ are swapped at $\epsilon_{43}=\Delta_{43}-\Omega_R$. Two of the four state combinations from the computational basis remain in $\left(1,2,1,2\right)$ at $\epsilon_{43}=\epsilon_{43}^*$. (b) shows the second step of the manipulation. $\epsilon_{23}=eV_2-eV_3$ models the potentials at $\text{QD}_2$ and $\text{QD}_3$. As a consequence, only $\ket{1_L0_R}$ and $\ket{0_L 0_R}$ can be tuned to $\left(1,1,2,2\right)$, but $\ket{1_L B}$ and $\ket{0_L B}$ remain in $\left(1,2,2,1\right)$. The nontrivial part of the entangling gate is a $\pi$-phase evolution of $\ket{0_L 0_R}$ at $\epsilon_{23}=\Delta_{23}$. $\ket{1_L 0_R}$ is degenerate with $\ket{L}$ and passes through a DLZC at $\epsilon_{23}=\Delta_{23}-\Omega_L$. Leakage from the computational subspace is prevented by the pulse cycle that involves waiting times at $\epsilon_{23}=\Delta_{23}-\Omega_L$ and at $\epsilon_{23}=\epsilon_{23}^*$ (see description in the text). The setup is brought back to the initial configuration in the end, by first changing $\epsilon_{23}$ and then changing $\epsilon_{43}$. Perfect state crossings are marked, where transitions are forbidden from spin-selection rules (blue), or from charge-selection rules (purple). The waiting times $t_1$, $t_2$, $t_w$, and $t_{\pi}$ are given in the text.
\label{fig:2}}
\end{figure}

One can construct an entangling operation in a two-step manipulation procedure, which is shown in \fref{fig:2}. In the first step, $\epsilon_{43}$ is modified, and the charge configuration is pulsed from $\left(1,2,1,2\right)$ towards $\left(1,2,2,1\right)$. Only $\ket{1_R}$ is transferred to $\ket{B}=\ket{\left(S\uparrow\right)_R}$ because $\ket{1_R}$ is energetically unfavored compared to $\ket{0_R}$, which remains in $\left(1,2\right)$. The tuning uses a rapid pulse to $\epsilon_{43}=\Delta_{43}-\Omega_R$. $\mathcal{H}_{34}$ couples $\ket{1_R}$ and $\ket{B}$ by $\sqrt{\frac{3}{2}}\tau_2$. The occupations of $\ket{1_R}$ and $\ket{B}$ swap after the waiting time $t_1=\frac{h}{2\sqrt{6}\tau_2}$. Afterwards, $\epsilon_{43}$ is pulsed to $\epsilon_{43}=\epsilon_{43}^*$, which is far away from all the anticrossings. $\ket{B}$ and $\ket{0_R}$ have the energy difference $\Omega_R^*$ at $\epsilon_{43}=\epsilon_{43}^*$. Note that $\epsilon_{43}=\epsilon_{43}^*$ is in the readout regime of $\text{HQ}_R$: $\ket{1_R}$ is in $\left(2,1\right)$, but $\ket{0_R}$ is in $\left(1,2\right)$.

In the second step, gate pulses modify $\epsilon_{23}$ at fixed $\epsilon_{43}=\epsilon_{43}^*$. The charge configuration is pulsed towards $\left(1,1,2,2\right)$. States in $\left(1,2,2,1\right)$ remain unchanged because they need the transfer of two electrons to reach $\left(1,1,2,2\right)$. The states
\begin{align}
\ket{L}=&\Big[\sqrt{\frac{1}{6}}\ket{\uparrow T_0 \uparrow}
-\frac{\sqrt{3}}{2}\ket{\uparrow T_+ \downarrow}
+\frac{1}{2\sqrt{3}}\ket{\downarrow T_+ \uparrow}\Big]\ket{S},\\
\ket{\beta}=&\frac{1}{2}\left[
\sqrt{2}\ket{\uparrow T_0 \uparrow}+
\ket{\uparrow T_+ \downarrow}+
\ket{\downarrow T_+ \uparrow}\right]\ket{S},
\end{align}
are introduced. $\ket{E}=\ket{\uparrow\uparrow S S}$ is the ground state in $\left(1,1,2,2\right)$ with $s_z=1$.
$\mathcal{H}_{23}$ couples $\ket{0_L0_R}$, $\ket{1_L0_R}$, $\ket{L}$, and $\ket{E}$, while $\ket{\beta}$ is decoupled. 
When approaching $\left(1,1,2,2\right)$, first the anticrossing of $\ket{1_L0_R}$, $\ket{L}$, and $\ket{E}$ is reached at $\epsilon_{23}=\Delta_{23}-\Omega_L$:
\begin{align}
\mathcal{H}_{23}\left(\epsilon_{23}\right)\approx
\left(
\begin{array}{ccc}
\Omega_L&0&\frac{\tau_4}{\sqrt{6}}\\
0&\Omega_L&-\frac{2\tau_4}{\sqrt{3}}\\
\frac{\tau_4}{\sqrt{6}}&-\frac{2\tau_4}{\sqrt{3}}&\Delta_{23}-\epsilon_{23}
\end{array}
\right).
\label{eq:HALL}
\end{align}
$\ket{0_L0_R}$ hybridizes with $\ket{E}$ only at $\epsilon_{23}=\Delta_{23}$. $\ket{E}$ has lower energy than $\ket{1_L0_R}$ at $\epsilon_{23}=\epsilon_{23}^*$, but $\ket{0_L0_R}$ is still the ground state.

The passage through the anticrossing at $\epsilon_{23}=\Delta_{23}-\Omega_L$ is critical for the construction of the entangling operation.
$\mathcal{H}_{23}$ describes within the subspace 
$\left\{\ket{1_L0_R},\ket{L},\ket{E}\right\}$ a DLZC (see \eref{eq:HALL}). A basis transformation partially diagonalizes \eref{eq:HALL}: $\ket{T_1}=\frac{1}{3}\ket{1_L0_R}-\frac{2\sqrt{2}}{3}\ket{L}$ and $\ket{E}$ have the overlap $\sqrt{3/2}\tau_4$, but $\ket{T_2}=\frac{2\sqrt{2}}{3}\ket{1_L0_R}+\frac{1}{3}\ket{L}$ is decoupled. $\ket{T_1}$ and $\ket{E}$ swap at $\epsilon_{23}=\Delta_{23}-\Omega_L$ after $t_{2}=\frac{h}{2\sqrt{6}\tau_4}$. One introduces the waiting time $t_{w}$ at $\epsilon_{23}=\epsilon_{23}^*$, where $\ket{E}$ has the energy $\Omega_L/2$. $t_{w}$ must compensate after the full cycle the relative phase evolution between $\ket{T_1}$ and $\ket{T_2}$; as a consequence, $\ket{1_L0_R}$ does not leak to $\ket{L}$. Simple mathematics shows that this is the case for $t_\text{w}=h\left(\frac{2n}{\Omega_L}-\frac{1}{\tau_3}\right)>0$ with $n\in\mathbb{N}$.

The time evolution at $\epsilon_{23}=\Delta_{23}$ constructs the central part of the entangling gate. $\mathcal{H}_{23}$ couples $\ket{0_L0_R}$ and $\ket{E}$ by $\tau_3$. The states of the subspace $\left\{\ket{0_L0_R},\ket{E}\right\}$ pick up a $\pi$-phase factor after the waiting time $t_\pi=\frac{h}{2\tau_3}$: $e^{-i\pi\sigma_x}=-\bm{1}$. All other states of the computational basis evolve trivially with the energies $\Omega_L$, $\Omega_R^*$, and $\Omega_L+\Omega_R^*$. Finally the setup is tuned back to the initial configuration, involving swaps at $\epsilon_{23}=\Delta_{23}-\Omega_L$ and $\epsilon_{43}=\Delta_{43}-\Omega_R$ that are generated after the waiting times $t_2=\frac{h}{2\sqrt{6}\tau_4}$ and $t_1=\frac{h}{2\sqrt{6}\tau_2}$.

In total, the described pulse cycle realizes a $\text{CPHASE}$ gate in the basis 
$\ket{1_L1_R}$,
$\ket{1_L0_R}$,
$\ket{0_L1_R}$, and 
$\ket{0_L0_R}$
when permitting additional single-qubit phase gates:
\begin{align}\label{eq:CPHASE}
&\mathcal{U}_{\epsilon_{43}=\Delta_{43}-\Omega_R}\left(t_1\right)
\mathcal{U}_{\epsilon_{23}=\Delta_{23}-\Omega_L}\left(t_2\right)
\mathcal{U}_{\epsilon_{23}=\Delta_{23}}\left(t_\pi\right)\\
&\nonumber\times\mathcal{U}_{\epsilon_{23}=\epsilon_{23}^*}\left(t_\text{w}\right)
\mathcal{U}_{\epsilon_{23}=\Delta_{23}-\Omega_L}\left(t_2\right)
\mathcal{U}_{\epsilon_{43}=\Delta_{43}-\Omega_R}\left(t_1\right)
\nonumber\\
&=e^{\frac{i\pi\left(p_1+p_2\right)}{2}}Z_L^{-\frac{p_1}{4}}Z_R^{-\frac{p_2}{4}}\text{CPHASE},
\nonumber
\end{align}
with $Z_i^{\phi}=e^{-i 2\pi \sigma_z^i\phi}$, 
$p_1=\Omega_R^*\left(\frac{1}{\tau_3}-\frac{2\sqrt{2/3}}{\tau_4}-\frac{4n}{\Omega_L}\right)$, and 
$p_2=\Omega_L\left(\frac{1}{\tau_3}-\frac{2\sqrt{2/3}}{\tau_4}\right)$. $\mathcal{U}_{\epsilon}\left(t\right)$ describes the time evolution at $\epsilon$ for the waiting time $t$. One has constructed a phase shift on $\text{HQ}_R$ conditioned on the state of $\text{HQ}_L$. 
\tref{Tab1} summarizes the manipulation steps of the CPHASE gate.

\begin{table*}
\begin{tabular}{ccccccccccr}
\hline
\hline
$\ket{1_L1_R}$&
\multirow{4}{*}{$\overset{(1)}{\longrightarrow}$}&
$\ket{1_LB}$&
\multirow{4}{*}{$\overset{(2)}{\longrightarrow}$}&
$\ket{1_LB}$&
\multirow{4}{*}{$\overset{(3)}{\longrightarrow}$}&
$\ket{1_LB}$&
\multirow{4}{*}{$\overset{(4)}{\longrightarrow}$}&
$\ket{1_LB}$&
\multirow{4}{*}{$\overset{(5)}{\longrightarrow}$}&
$\ket{1_L1_R}$\\
$\ket{1_L0_R}$&&$\ket{1_L0_R}$&&
$\left\{\ket{1_L0_R},\ket{L},\ket{E}\right\}$&&
$\left\{\ket{1_L0_R},\ket{L},\ket{E}\right\}$&&
$\ket{1_L0_R}$&&$\ket{1_L0_R}$\\
$\ket{0_L1_R}$&&$\ket{0_LB}$&&$\ket{0_LB}$&&$\ket{0_LB}$&&$\ket{0_LB}$&&$\ket{0_L1_R}$\\
$\ket{0_L0_R}$&&$\ket{0_L0_R}$&&$\ket{0_L0_R}$&&$-\ket{0_L0_R}$&&$-\ket{0_L0_R}$&&$-\ket{0_L0_R}$\\
\hline
\hline
\end{tabular}
\caption{
\label{Tab1}
Summary of the state evolution that generates the CPHASE gate, as described in the text [see \eref{eq:CPHASE}]. All phase evolutions that can be corrected with single-qubit gates are neglected. (1) The states $\ket{1_R}$ and $\ket{B}$ interchange. (2) The transfer through a DLZC mixes the state $\ket{1_L0_R}$ to the subspace $\left\{\ket{1_L0_R},\ket{L},\ket{E}\right\}$. (3) The central part of the entangling operation introduces a nontrivial phase factor to $\ket{0_L0_R}$. (4) The content in $\left\{\ket{1_L0_R},\ket{L},\ket{E}\right\}$ is brought back to $\ket{1_L0_R}$ using the appropriate pulse shape. (5) $\ket{B}$ and $\ket{1_R}$ interchange.
}
\end{table*}

\section{Gate Performance and Noise Properties
\label{sec:Noise}}
In general, two-qubit pulse gates are fast. The only time consuming parts of the entangling gate are the waiting times at $\epsilon_{43}=\Delta_{43}-\Omega_R$, $\epsilon_{23}=\Delta_{23}-\Omega_L$, $\epsilon_{23}=\epsilon_{23}^*$, and $\epsilon_{23}=\Delta_{23}$. The overall gate time is on the order of $\mathcal{O}\left(\frac{h}{\tau_2},\frac{h}{\tau_3},\frac{h}{\tau_4}\right)$. It was shown that tunnel couplings between QDs of a DQD in Si reach $3~\mu\text{eV}$ \cite{wu2014,maune2012}. Two DQDs might be some distance apart from each other; nevertheless, $\mu\text{eV}$ tunnel couplings seem possible. An entangling gate will take only a few nanoseconds but requires subnanosecond pulses.

\begin{figure}[htb]
\centering
\includegraphics[width=0.49\textwidth]{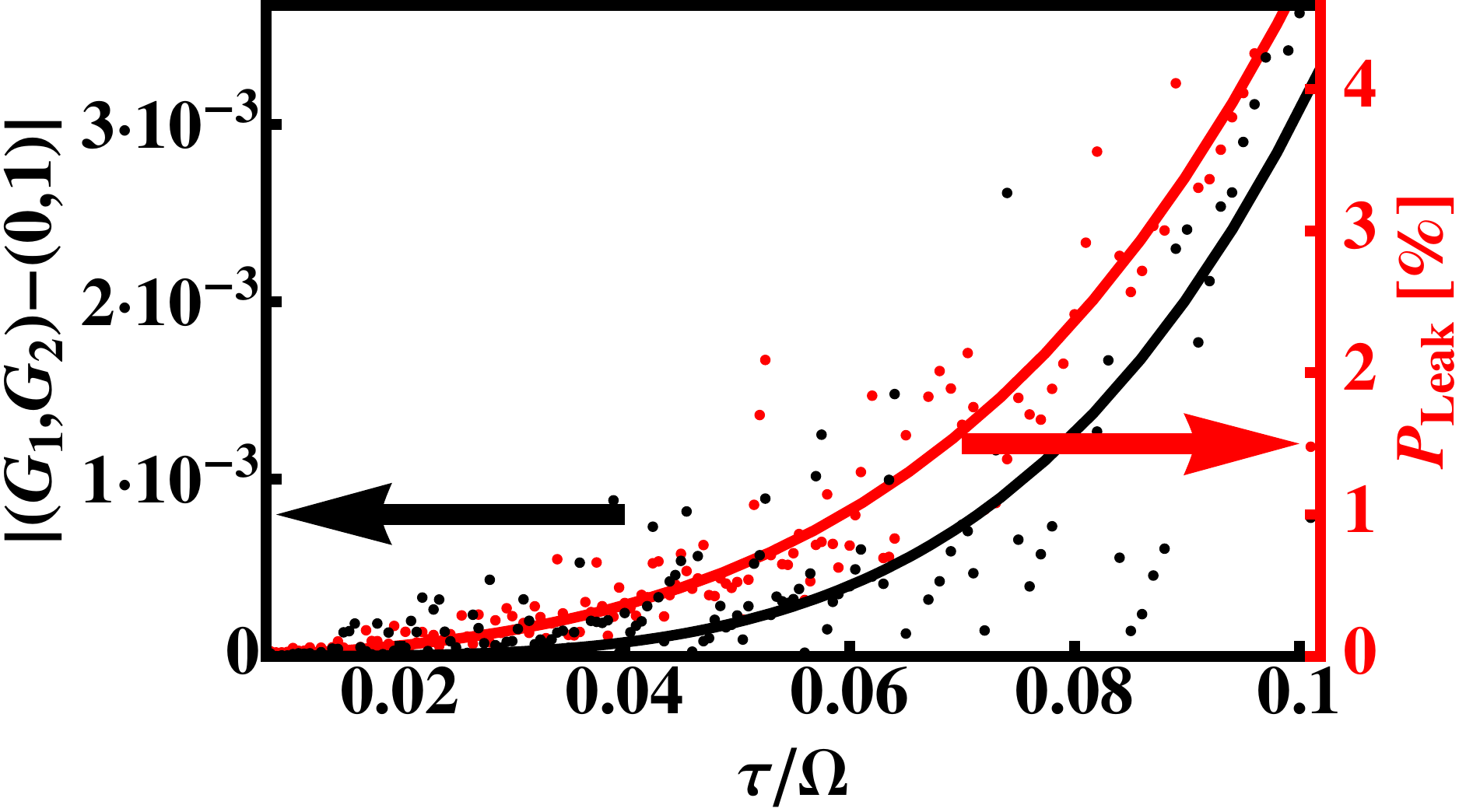}
\caption{Numerically optimized gate sequences according to \eref{eq:CPHASE} for $\Omega=\Omega_L=\Omega_R$ and $\tau=\tau_i$, $i=1,\dots,4$. The deviations of the Makhlin invariants \cite{makhlin2002} from $G_1=0$ and $G_2=1$ and the leakage errors $P_{\text{Leak}}$ are numerically minimized by adjusting the waiting times and waiting positions. $P_{\text{Leak}}=\left|U_{\mathcal{PQ}}\right|^2$ is the transition probability from the computational subspace $\mathcal{P}$ to the leakage subspace $\mathcal{Q}$. The points describe single numerical results; the solid lines are a polynomial fit. Note that small $\tau/\Omega$ permit better gates.
\label{fig:3}}
\end{figure}

The setup provides a rich variety of leakage states. \aref{app:EXTBASIS} introduces an extended state basis in $s_z=1$. I consider the charge configurations $\left(1,2,1,2\right)$, $\left(1,2,2,1\right)$, and $\left(1,1,2,2\right)$, while I neglect doubly occupied triplets at $\text{QD}_1$ and $\text{QD}_3$ (see \sref{sec:Setup}). The tunnel couplings are only relevant around state degeneracies in the gate construction, which is justified for vanishing $\tau_i$, $i=1,\dots,4$, compared to $\Omega_L$ and $\Omega_R$. In reality, $\tau_i$ are small compared to $\Omega_L$ and $\Omega_R$, but they are not negligible. As a consequence, modifications from the anticrossings partially lift the neighboring state crossings (see the blue and purple circles in \fref{fig:2}) and modify the energy levels and anticrossings. \fref{fig:3} shows that high-fidelity gates can be constructed that only have small leakage, when the waiting times and the waiting positions introduced earlier are adjusted numerically. Small leakage errors and minor deviations from a $\text{CPHASE}$ gate are reached for $\tau_{i}/\Omega_{L,R}<5\%$, $i=1,\dots,4$. I use $\Omega/h=\Omega_{L}/h=\Omega_R/h=15\ \text{GHz}$ and $\tau/h=\tau_{i}/h=0.5\ \text{GHz}$, $i=1,\dots,4$ in the following noise analysis (see \rcite{mehl2014} for a similar noise discussion).

\subsection{Charge Noise}
Charge traps of the heterostructure introduce low-frequency electric field fluctuations \cite{petersson2010,dial2013}. Their influence is weak for spin qubits, but it increases for charge qubits \cite{hu2006,mehl2013}. Consequently, HQs are protected from charge noise only in the idle configuration. Charge noise is modeled by a low-frequency energy fluctuation between different charge configurations. I introduce no fluctuations during one gate simulation, but use modifications between successive runs. The fluctuations follow a Gaussian probability distribution of rms $\delta \epsilon$. Note that the numerically optimized gate sequence of \eref{eq:CPHASE} is simulated.

\fref{fig:4} shows the gate fidelity $F$, which is defined in \aref{app:FID}, while $\delta \epsilon$ is varied. $F$ decreases rapidly with $\delta \epsilon$. A Gaussian decay is seen for small $\delta \epsilon$. The decay constant shows that $\tau$ is the relevant energy scale of the entangling gate. The coherence is lost if $\delta \epsilon$ increases beyond $\tau$ because a typical gate misses the anticrossings of \fref{fig:2}. Noisy gate sequences keep only the diagonal entries of the density matrix, but they remove all off-diagonal entries leading to $F=0.25$.

Charge noise can be modeled for QD spin qubits to cause energy fluctuations of $\delta \epsilon \approx \mu\text{eV}$ ($1~\mu\text{eV}/h\approx 0.2~\text{GHz}$). Both for GaA charge qubits \cite{petersson2010} and Si charge qubits \cite{shi2013}, current experiments suggest charge noise on the order of a few $\mu\text{eV}$.
For high-fidelity pulse-gated entangling operations, $\delta \epsilon$ must be smaller than $\tau$ that reaches typically a few $\mu\text{eV}$ in Si HQs.

\subsection{Hyperfine Interactions}
Nuclear spins couple to HQs, and they cause low-frequency magnetic field fluctuations \cite{taylor2007,hanson2007-2}.
The error analysis can be restricted to the total $s_z=1$ subspace when the global magnetic fields $E_z$ are larger than the uncertainties in the magnetic field $\delta E_z$ at every QD. Already global magnetic fields of $100~\text{mT}$ are much larger than the typical $\delta E_z$ for Si QDs [$E_z/h>3~\text{GHz}$ ($>100~\text{mT}$) and $\delta E_z/h<3~\text{MHz}$ ($<100~\mu\text{T}$) for Si QDs \cite{assali2011}$^{,}$\footnote{
Also GaAs QDs would fulfill this condition with $E_z/h>0.5~\text{GHz}$ ($>100~\text{mT}$) and $\delta E_z/h<30~\text{MHz}$ ($<5~\text{mT}$), which describes an uncorrected nuclear spin bath.
}].
I simulate the numerically optimized pulse sequence of \eref{eq:CPHASE} under magnetic field fluctuations. The variations of the magnetic fields at every QD are determined by a Gaussian probability distribution with the rms $\delta E_z$ (in energy units).

\begin{figure}[htb]
\centering
\includegraphics[width=0.49\textwidth]{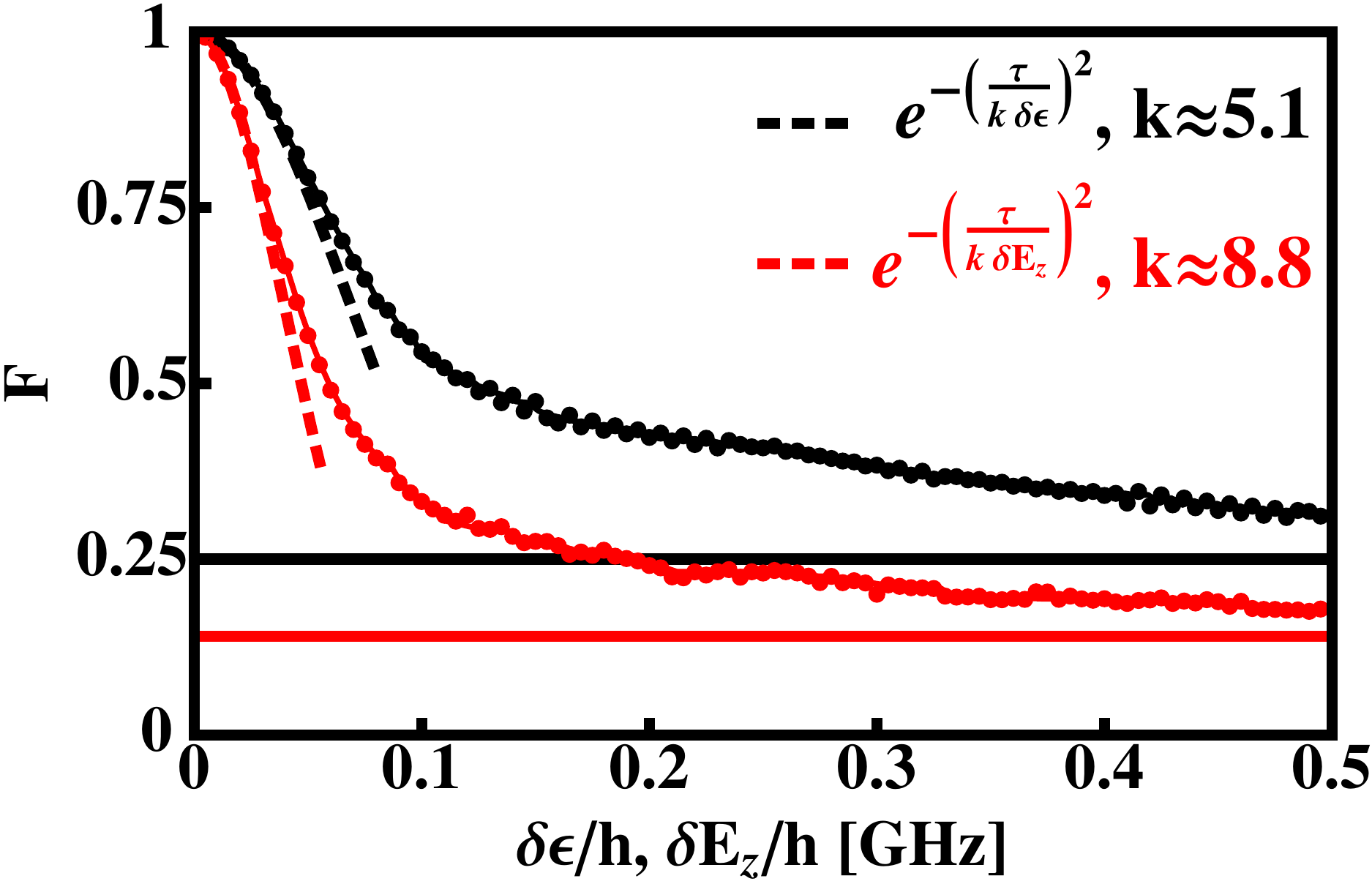}
\caption{Fidelity analysis for the numerically optimized $\text{CPHASE}$ gates under charge noise (black) and nuclear spin noise (red) at $\Omega_{L}/h=\Omega_R/h=15\ \text{GHz}$ and $\tau/h=\tau_{i}/h=0.5\ \text{GHz}$, $i=1,\dots,4$. The energy fluctuations $\delta \epsilon$ between different charge configurations model charge noise. Nuclear spins cause local, low-frequency magnetic field fluctuations of the energy $\delta E_z$. Both noise sources can be described by a classical probability distribution with the rms $\delta \epsilon$ (for charge noise) and $\delta E_z$ (for nuclear spin noise). The fidelity $F$ is extracted from $1000$ gate simulations according to \eref{eq:CPHASE}. Increasing the uncertainties suppresses $F$ strongly till it saturates at $0.25$ (for charge noise) and $9/64$ (for nuclear spin noise) (see the horizontal lines). The initial decay of $F$ is described by a Gaussian decay law (see the dotted lines).
\label{fig:4}}
\end{figure}

\fref{fig:4} shows that $F$ decreases rapidly with $\delta E_z$. Again, a Gaussian decay is observed with a decay constant determined by $\tau$ for small $\delta E_z$. The influence of hyperfine interactions differs from charge noise. Local magnetic fields lift the state crossings that are protected by the spin-selection rules (see blue markings in \fref{fig:2}). Not only is the coherence lost for large $\delta E_z$, but leakage further suppresses $F$. The limit of large $\delta E_z$ can be approximated with $F=9/64$. All off-diagonal entries of the density matrix are removed. Additionally, some states are mixed with leakage states. $\ket{1_L1_R}$ goes to a mixed state with three other states; $\ket{1_L0_R}$ and $\ket{0_L1_R}$ mix with one other state each.

Si is a popular QD material because the number of finite-spin nuclei is small \cite{zwanenburg2013}. Nevertheless, noise from nuclear spins was identified to be dominant in the first spin qubit manipulations of gate-defined Si QDs \cite{maune2012}. $\delta E_z/h=7.5\cdot 10^{-4}~\text{GHz}$ in natural Si (see \rcite{assali2011}) is sufficient for nearly perfect two-qubit pulse gates. The fluctuations of the nuclear spins decrease further for isotopically purified Si instead of natural Si,
a system which has shown rapid experimental progress recently \cite{veldhorst2014,veldhorst2014-2}. We note that $\delta E_z/h=30~\text{MHz}$ for GaAs QDs would be problematic for high-fidelity entangling operations.

\section{Conclusion
\label{sec:Concl}}
I have constructed a two-qubit pulse gate for the HQ --- a qubit encoded in a three-electron configuration on a gate-defined DQD. Applying fast voltage pulses at gates close to the QDs enables the transfer of single electrons between QDs. The setup is tuned to the anticrossing of $\ket{0_L 0_R}$ with the leakage state $\ket{E}$. $\ket{0_L 0_R}$ picks up a nontrivial phase without leaking to $\ket{E}$, while all the other two-qubit states accumulate trivial phases. The main challenge of the entangling gate is to avoid leakage to other states. One can use a two-step procedure. (1) The right HQ is pulsed to the readout configuration. Here, $\ket{1_R}$ goes to $\left(2,1\right)$, but $\ket{0_R}$ stays in $\left(1,2\right)$. (2) $\ket{0_L 1_R}$ passes through a DLZC during the pulse cycle. The pulse profile is adjusted to avoid leakage after the full pulse cycle. Note that an adiabatic manipulation protocol can substitute the pulse-gated manipulation\footnote{
All energy levels follow the lowest energy states for adiabatic manipulation protocols. The nontrivial part of the entangling gate is also obtained at the degeneracy of $\protect{\ket{0_L0_R}}$ with $\protect{\ket{E}}$. The pulse shape must compensate for the pulsing through the DLZC of $\protect{\left\{\ket{1_L0_R},\ket{L},\ket{E}\right\}}$.
}.

Cross-couplings between anticrossings, charge noise, and nuclear spin noise introduce errors for the pulse-gated two-qubit operation. Cross-couplings between anticrossings are problematic as they open state crossings. Also these mechanism slightly influence the energy levels and the sizes of the anticrossings. Reasonably small values of $\tau/\Omega\lesssim 5\%$ still permit excellent gates through pulse shaping. Charge noise is problematic because the gate tunes the HQs between different charge configurations. Current QD experiments suggest that charge noise is critical for the pulse-gated entangling operation. Nuclear spins are unimportant for the pulse-gated entangling operation of HQs in natural Si and, even more, for isotopically purified Si. I am hopeful that material improvements and advances in fabrication techniques for Si QDs still allow an experimental realization of this gate in the near future. 

Pulse gates provide universal control of HQs through single-qubit operations, which have been implemented experimentally \cite{shi2014,kim2014}, together with the described two-qubit entangling gate. Because this setup can be scaled up trivially (see \fref{fig:1}), further experimental progress should be stimulated to realize all-pulse-gated manipulations of HQs.

\textit{Acknowledgments} \textthreequartersemdash\ 
I thank D. P. DiVincenzo and L. R. Schreiber for many useful discussions.

\begin{appendix}
\section{Fidelity Description of Noisy Gates
\label{app:FID}}
$U_{n}^{\xi}$ describes a noisy operation with a parameter $\xi$ which modifies the gate between different runs of the experiment and obeys a classical probability distribution $f\left(\xi\right)$. The entanglement fidelity is a measure for the gate performance \cite{nielsen2000,marinescu2012}: 
\begin{align}
F\left(\xi\right)=\text{tr}
\left\{\rho^{RS}
\bm{1}_R\otimes\left[U_{i}^{-1}U_{n}^{\xi}\right]_S
\rho^{RS}
\bm{1}_R\otimes\left[(U_{n}^{\xi})^{-1}U_{i}\right]_S
\right\}.
\label{eq:FIDDEF}
\end{align}
$U_{i}$ describes the ideal time evolution. The state space is doubled to two identical Hilbert spaces $R$ and $S$. $\rho^{RS}=\op{\psi}{\psi}$ is a maximally entangled state on the larger Hilbert space; e.g., $\ket{\psi}=\left(\ket{0000}+\ket{0110}+\ket{1001}+\ket{1111}\right)/2$. The gate fidelity $F$ is calculated by averaging \eref{eq:FIDDEF} over many instances of $U_{n}^\xi$, giving $F=\int d\xi~f\left(\xi\right) F\left(\xi\right)$. $F=1$ for perfect gates. This definition captures also leakage errors.

\section{Extended Basis
\label{app:EXTBASIS}}
\tref{Tab2} provides an extended state basis in $s_z=1$ for the description of two HQs in $\left(1,2,1,2\right)$, $\left(1,2,2,1\right)$, and $\left(1,1,2,2\right)$. States with a doubly occupied triplet at $\text{QD}_1$ or $\text{QD}_3$ are neglected because the triplet configurations at $\text{QD}_1$ and $\text{QD}_3$ are assumed to require much higher energies than the singlet configurations (see \sref{sec:Setup}). $\ket{1_L1_R}$, $\ket{1_L0_R}$, $\ket{0_L1_R}$, and $\ket{0_L0_R}$ are the computational basis of two HQs. The states $\ket{L}$, $\ket{1_LB}$, and $\ket{0_LB}$ are partially filled during the manipulation procedure. All other states are leakage states that are ideally unfilled during the manipulation. The states describe the spin configurations at $\text{QD}_i$, $i=1,\dots,4$, of the array of four QDs, and they are grouped into subspaces of equal energies.

It is straight forward to prove that the $23$ states in \tref{Tab2} are a complete set to describe the six-electron spin problem of two HQs. Note that the discussion is restricted to total $s_z=1$. One needs two additional spin-$\uparrow$ electrons compared to the spin-$\downarrow$ electrons in the $\left(1,2,1,2\right)$ configuration, giving in total $\left(\begin{array}{c}6\\4\end{array}\right)=15$ choices. In the $\left(1,2,2,1\right)$ and $\left(1,1,2,2\right)$ configurations, the electrons at $\text{QD}_2$ and at $\text{QD}_4$ are always paired to a singlet state (because it is strongly unfavored to reach a triplet at these QDs), giving $\left(\begin{array}{c}4\\3\end{array}\right)=4$ choices to reach in total $s_z=1$.

\begin{table*}
\begin{center}
\begin{tabular}{lll}
\hline
\hline
& \multicolumn{1}{c}{state} & \multicolumn{1}{c}{\ \ energy\ \ \ \ \ }\\
\hline

\parbox[t]{2mm}{\multirow{14}{*}{\rotatebox[origin=c]{90}{
-----------------------------------
(1,2,1,2)
-----------------------------------
}}}
& $\ket{1_L1_R}=
\left[\sqrt{\frac{2}{3}}\ket{\downarrow T_+}-\sqrt{\frac{1}{3}}\ket{\uparrow T_0}\right]
\left[\sqrt{\frac{2}{3}}\ket{\downarrow T_+}-\sqrt{\frac{1}{3}}\ket{\uparrow T_0}\right]$
& \rdelim\}{9}{4mm}[$\Omega_L+\Omega_R$]\\
& $\ket{\alpha_1}=
\left[\sqrt{\frac{1}{3}}\ket{\downarrow T_+}+\sqrt{\frac{2}{3}}\ket{\uparrow T_0}\right]
\left[\sqrt{\frac{2}{3}}\ket{\downarrow T_+}-\sqrt{\frac{1}{3}}\ket{\uparrow T_0}\right]$\\
& $\ket{\alpha_2}=
\left[\sqrt{\frac{2}{3}}\ket{\downarrow T_+}-\sqrt{\frac{1}{3}}\ket{\uparrow T_0}\right]
\left[\sqrt{\frac{1}{3}}\ket{\downarrow T_+}+\sqrt{\frac{2}{3}}\ket{\uparrow T_0}\right]$\\
& $\ket{\alpha_3}=
\left[\sqrt{\frac{1}{3}}\ket{\downarrow T_+}+\sqrt{\frac{2}{3}}\ket{\uparrow T_0}\right]
\left[\sqrt{\frac{1}{3}}\ket{\downarrow T_+}+\sqrt{\frac{2}{3}}\ket{\uparrow T_0}\right]$\\
& $\ket{\alpha_4}=\ket{\uparrow T_- \uparrow T_+}$\\
& $\ket{\alpha_5}=\ket{\uparrow T_+ \uparrow T_-}$\\
& $\ket{\alpha_6}=\ket{\uparrow T_+ \downarrow T_0}$\\
& $\ket{\alpha_7}=\ket{\downarrow T_0 \uparrow T_+}$\\
\\
& $\ket{1_L0_R}=
\left[\sqrt{\frac{2}{3}}\ket{\downarrow T_+}-\sqrt{\frac{1}{3}}\ket{\uparrow T_0}\right]\ket{\uparrow S}$
& \rdelim\}{4}{4mm}[$\Omega_L$]\\
& $\ket{L}=\left[
\sqrt{\frac{1}{6}}\ket{\uparrow T_0 \uparrow}
-\frac{\sqrt{3}}{2}\ket{\uparrow T_+ \downarrow}
+\frac{1}{2\sqrt{3}}\ket{\downarrow T_+ \uparrow}
\right]\ket{S}$\\
& $\ket{\beta}=\left[
\frac{1}{2}\ket{\uparrow T_+\downarrow}
+\frac{1}{2}\ket{\downarrow T_+\uparrow}
+\sqrt{\frac{1}{2}}
\ket{\uparrow T_0\uparrow}
\right]\ket{S}$\\
\\
& $\ket{0_L1_R}=
\ket{\uparrow S}\left[\sqrt{\frac{2}{3}}\ket{\downarrow T_+}-\sqrt{\frac{1}{3}}\ket{\uparrow T_0}\right]$
& \rdelim\}{3}{4mm}[$\Omega_R$]\\
& $\ket{\gamma_1}=
\ket{\uparrow S}\left[\sqrt{\frac{1}{3}}\ket{\downarrow T_+}+\sqrt{\frac{2}{3}}\ket{\uparrow T_0}\right]$\\
& $\ket{\gamma_2}=
\ket{\downarrow S \uparrow T_+}$\\
\\
& $\ket{0_L0_R}=
\ket{\uparrow S \uparrow S}$& $\ \ 0$\\
\\
\parbox[t]{2mm}{\multirow{4}{*}{\rotatebox[origin=c]{90}{
---
(1,2,2,1)
---
}}}
& $\ket{1_LB}=
\left[\sqrt{\frac{2}{3}}\ket{\downarrow T_+}-\sqrt{\frac{1}{3}}\ket{\uparrow T_0}\right]\ket{S\uparrow}$
& \rdelim\}{3}{4mm}[$\Delta_{43}+\Omega_L$]\\
& $\ket{\delta_1}=
\left[\sqrt{\frac{1}{3}}\ket{\downarrow T_+}+\sqrt{\frac{2}{3}}\ket{\uparrow T_0}\right]\ket{S\uparrow}$\\
& $\ket{\delta_2}=
\ket{\uparrow T_+ S\downarrow}$\\
\\
& $\ket{0_LB}=
\ket{\uparrow SS\uparrow}$& $\ \ \Delta_{43}$\\
\\
\parbox[t]{2mm}{\multirow{4}{*}{\rotatebox[origin=c]{90}{
--
(1,1,2,2)
--
}}}
& $\ket{\mu_1}=\ket{\uparrow\uparrow ST_0}$& \rdelim\}{3}{4mm}[$\Delta_{23}+\Omega_R$]\\
& $\ket{\mu_2}=\ket{\uparrow\downarrow ST_+}$\\
& $\ket{\mu_3}=\ket{\downarrow\uparrow ST_+}$\\
\\
& $\ket{E}=\ket{\uparrow\uparrow SS}$& $\ \ \Delta_{23}$\\
\hline
\hline
\end{tabular}
\caption{Extended state basis with the total spin quantum number $s_z=1$ for the setup of six electrons distributed over four QDs. Each entry of the states describes a spin configuration at one of the QDs with the notation $\ket{\text{QD}_1,\text{QD}_2,\text{QD}_3,\text{QD}_4}$. All the relevant states for the electron configurations $\left(n_{\text{QD}_1},n_{\text{QD}_2},n_{\text{QD}_3},n_{\text{QD}_4}\right)=\left(1,2,1,2\right)$, $\left(1,2,2,1\right)$, and $\left(1,1,2,2\right)$ are included. Further details are given in the text.
\label{Tab2}}
\end{center}
\end{table*}

\end{appendix}

\bibliography{library}
\end{document}